\documentclass{emulateapj}
\usepackage{graphicx}
\begin{document} 

% ApJ Title
% ApJ Title
\title{The Influence of Orbital Eccentricity on Tidal Radii of Star Clusters}
\author{Jeremy J. Webb$^1$, William E. Harris$^1$, Alison Sills$^1$, Jarrod R. Hurley$^2$}
\affil{$^1$ Department of Physics and Astronomy, McMaster University, Hamilton ON L8S 4M1, Canada}
\affil{$^2$ Centre for Astrophysics and Supercomputing, Swinburne University of Technology, P.O. Box 218, VIC 3122, Australia}
\email{webbjj@mcmaster.ca}

%************************************ABSTRACT******************************************
\begin{abstract}

We have performed $N$-body simulations of star clusters orbiting in a spherically symmetric smooth galactic potential. The model clusters cover a range of initial half-mass radii and orbital eccentricities in order to test the historical assumption that the tidal radius of a cluster is imposed at perigalacticon. The traditional assumption for globular clusters is that since the internal relaxation time is larger than its orbital period, the cluster is tidally stripped at perigalacticon. Instead, our simulations show that a cluster with an eccentric orbit does not need to fully relax in order to expand. After a perigalactic pass, a cluster re-captures previously unbound stars, and the tidal shock at perigalacticon has the effect of energizing inner region stars to larger orbits. Therefore, instead of the limiting radius being imposed at perigalacticon, it more nearly traces the instantaneous tidal radius of the cluster at any point in the orbit. We present a numerical correction factor to theoretical tidal radii calculated at perigalacticon which takes into consideration both the orbital eccentricity and current orbital phase of the cluster. 

\end{abstract}

%************************************INTRODUCTION************************************

\section{Introduction \label{Introduction}}

Theoretical calculations of the radius of a star cluster, or \textit{tidal radius}, usually assume that the gravitational field of the host galaxy regulates cluster size. In most previous treatments, for simplicity it is further assumed that the gravitational field in which the cluster orbits is constant, i.e. the cluster has a circular orbit in a spherically symmetric galactic potential \citep[e.g.][]{vonhoerner57, king62, innanen83, jordan05, binney08, bertin08}. The tidal radius is then assumed to be equal to the \textit{Jacobi radius} ($r_J$), the distance beyond which the acceleration a star feels towards the galaxy center is greater than the acceleration it feels towards the cluster center, and the star is able to escape. First-order tidal theory determines the tidal radius \citep{vonhoerner57} via:

\begin{equation}\label{rtHoerner}
r_t=R_{gc}(\frac{M_{cl}}{2M_g})^{1/3}
\end{equation}

\noindent where $R_{gc}$ is the galactocentric distance of the cluster, $M_{cl}$ is cluster mass, and $M_g$ is the mass of the galaxy (assumed in early studies to be a point mass). 

For a cluster with a non-circular orbit, the fact that the tidal field is no longer static makes an analytic approach very difficult (see \citet{renaud11} for another approach). Historically it has been assumed that for a globular cluster on an eccentric orbit, its tidal radius is imposed at perigalacticon ($R_p$) where the tidal field of the host galaxy is the strongest. This assumption was initially suggested by \citet{vonhoerner57} and later \citet{king62}, and follows from the fact that the internal relaxation time ($t_{r_h}$) of the cluster is greater than its orbital period for almost all observed globular clusters. Therefore after stars outside the tidal radius at perigalacticon escape, the cluster would not be able to relax and expand before it returns to perigalacticon. Thus in Equation \ref{rtHoerner}, $R_{gc}$ is usually replaced with $R_p$ to calculate the tidal radius of a cluster with an eccentric orbit \citep[e.g.][]{innanen83, fall01, read06, webb12}.

However, recent studies are showing with increasingly strong evidence that the actual sizes of observed clusters are not imposed at perigalacticon in this simple way. The actual size of an observed cluster is known as its limiting radius $r_L$, which marks the point where the cluster density falls to zero \citep{binney08}. Using the solved orbits of 15 Galactic globular clusters, \citet{odenkirchen97} demonstrated that cluster limiting radii are not solely dependent on perigalactic distance. \citet{brosche99} suggested some sort of orbit-averaged tidal radius is more appropriate when predicting limiting radii. Even with the orbits of an additional 29 Milky Way globular clusters currently known \citep{dinescu99, dinescu07}, there is still no clear relationship between limiting radii and perigalactic distance, and the conclusions of \citet{odenkirchen97} still hold.

\begin{figure}[tbp]
\centering
\includegraphics[width=\columnwidth]{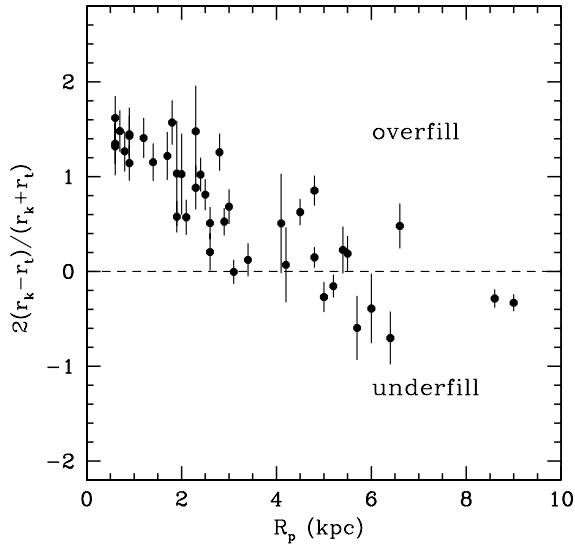}
\caption{Ratio of difference between observed ($r_k$) and theoretical ($r_t$) tidal radius at perigalacticon to the average ($(r_t+r_k)/2$) versus perigalactic distance for Galactic globular clusters.}
  \label{fig:rt_rp_norm}
\end{figure}

With the Galactic potential as given by \citet{johnston95} (the same Galactic potential Casetti-Dinescu et al. used to solve cluster orbits), we calculated the \textit{theoretical} tidal radius $r_t$ at perigalacticon (that is, the Jacobi radius at $R = R_p$) of each of the 44 Galactic clusters with known orbits. The $r_t$ values were calculated with the formalism of \citet{bertin08} (as outlined in Section 3.0). The main uncertainty in the theoretical tidal radius is due to the uncertainty in $R_p$ quoted in \citet{dinescu99} and \citet{dinescu07}. To compare theory and observations for individual clusters, we take cluster limiting radii as determined from direct \citet{king66} model fits $r_k$ to the observed cluster profiles as listed in \citet{harris96} (2010 edition). We then calculate the ratio of the difference ($r_k-r_t$) between observed and theoretically predicted values to their average ($(r_k+r_t)/2$). An uncertainty of $10 \%$ was assigned to values of $r_k$. If theory and observations are in agreement, the ratio will be approximately zero. Clusters which have a ratio greater than zero will be ones which overfill their predicted tidal radius, while clusters with ratios less than zero will be tidally under-filling. The comparison between theory and observations is illustrated in Figure \ref{fig:rt_rp_norm}. 

While it is not expected that all clusters are tidally filling \citep{gieles10}, the fact that the majority of clusters appear to be tidally overfilling is a strong signal that the simple assumptions built into the model need investigation. The known presence of tidal tails around observed \citep[e.g.][]{odenkirchen01} and simulated \citep[e.g.][]{montuori07, kupper12, lane12} globular clusters is not sufficient to explain the cases of apparent overfilling. The observed $r_k$ values are determined from King-model profile fits that are heavily dominated by the inner populations of stars, out to a few half-light radii. In almost all cases the extremely low densities of stars in the tails, at or beyond the nominal tidal radius, exert little leverage on these fits. Furthermore, \cite{king66} models are known to underestimate cluster sizes in general as they require a sharp tidal cutoff which is not observed in all clusters \citep{mclaughlin05}. Only in the most extreme cases (e.g. Pal 5 \citep{odenkirchen03}), will large and extended tidal tails influence model fits to the cluster surface brightness profile.

Theoretical calculations may also underestimate $r_k$ because the assumption that the tidal radius is imposed at perigalacticon implies that the shape of the tidal field and the cluster orbit are not important. Hence factors such as tidal heating and disk shocking due to a varying symmetric galactic potential are not taken into consideration. However, despite these potential inaccuracies we still expect to see some sort of correlation between $r_k$ and perigalactic distance, if tidal radii are imposed at perigalacticon. The Milky Way cluster data therefore suggest that something is wrong with basic tidal theory. 
 
Recent $N$-body simulations by \citet{kupper10}, find that their fitted \citet{king62} radius was better represented by the time averaged mean tidal radius of the cluster and not the perigalactic tidal radius. In a later study on the structure of tidal tails, \citet{kupper12} found that while stars outside the tidal radius as calculated at perigalacticon will likely become unbound at perigalacticon, some are able to be re-captured by the cluster as it moves away from perigalacticon and the instantaneous tidal radius of the cluster increases. That is, the limiting radius of a cluster will be greater than the tidal radius calculated at perigalacticon. This discrepancy is expected to be amplified for clusters on very eccentric orbits, as they make quick perigalactic passes and spend the majority of their time near apogalacticon ($R_{a}$). In fact, $N$-body simulations by \citet{madrid12} suggest that the half-mass radius of a globular cluster is more likely imposed at $R_a$ than $R_p$.

The purpose of this study is to explore more thoroughly the influence of orbital eccentricity on cluster size. Model $N$-body clusters with different initial half-mass radii are evolved from zero age to 10 Gyr over a range of orbital eccentricities in the disk of a Milky Way-like potential. The models and their initial conditions are described in Section 2. In Section 3 we focus on the influence of orbital eccentricity on the mass ($M$), half-mass radius ($r_m$), limiting radius, and tidal radius of each model cluster over time. In Section 4 we explore the influence of initial cluster half-mass radius on our results. In Section 5 we discuss the results of all our $N$-body models and the influence of orbital eccentricity not only on cluster radii, but on individual stars within the cluster as well. Based on our findings, in Section 6 we suggest a correction factor that can be applied to the perigalactic tidal radius of a cluster to better match its observed limiting radius. This correction factor is then applied to the Galactic globular clusters shown in Figure \ref{fig:rt_rp_norm}. Finally in Section 7 we summarize our conclusions.

\section{The Models \label{stwo}}

We use the NBODY6 direct $N$-body code \citep{aarseth03} to evolve model star clusters from zero age to a Hubble time, over a range of both initial cluster half-mass radii and orbital eccentricity. All models in this study begin with 48000 single stars and 2000 binaries.

As long as we use one particle to represent one real star, clusters of this size correspond physically to either a very massive open cluster, or a low-mass globular cluster in the Milky Way. Ideally, we would like to follow more ``average'' globular clusters, which will typically have 200,000 stars or more. However, to this point in the history of the subject, direct N-body integrations with N-values that large have only been carried out for special, specific purposes (for example, see \citet{hurley12} for a 200,000-star simulation in the tidal field of a point-mass Galactic potential and a circular orbit; \citet{zonoozi11} for $N \leq 100,000-$star simulations directed at modelling Pal 14, again on a circular orbit; or \citet{heggie09} for a 105,000-star simulation of NGC 6397 over 1 Gyr). Encouragingly, however, these high-N models generally confirm the trends obtained from earlier, small-N simulations (see Hurley et al. for additional discussion).

The purpose of our suite of 50,000-star models is instead to survey a wide parameter space of initial cluster half-mass radii and orbital type. Our chosen ranges (see below) match real Milky Way star clusters moving in a realistic time-varying potential. Eventually, with advances in computational capabilities this type of survey work can be extended to higher N values (up to $10^6$) that will cover almost the entire known range of globular clusters. However, as will be seen below, the models summarized here already prove to be highly informative in revealing important physical effects of orbital eccentricity on the internal dynamical evolution in a direct way that does not rely on analytical approximations.

Since we are only concerned with the influence of orbital eccentricity on clusters of different initial half mass radii, our choice of initial parameters such as cluster metallicity, the stellar initial mass function (IMF), and binary fraction are of little consequence as long as they remain consistent between models. However, we note that binary fractions of a few per cent are typical for globular clusters \citep[e.g.][]{davis08}.

The masses of single stars are drawn from a \citet{kroupa93} IMF between 0.1 and 30 $M_{\odot}$. For binary stars, the masses of two randomly selected single stars are combined to equal the total mass of the binary, with the primary and secondary masses determined by a mass-ratio randomly drawn from a uniform distribution. The initial total mass of each model is $3 \times 10^4 M_\odot$. The initial period of each binary is drawn from the distribution of \citet{duquennoy91} and their orbital eccentricities are assumed to follow a thermal distribution \citep{heggie75}. All stars were given a metallicity of $Z=0.001$. The initial positions and velocities of the stars are generated based on a Plummer density profile \citep{plummer11,aarseth74}. We note that the Plummer model extends to an infinite radius so we impose a cut-off at $\sim 10 \ r_m$ to avoid rare cases of large cluster-centric distances. A description of the algorithms for stellar and binary evolution can be found in \citet{hurley08a, hurley08b}.

The model clusters orbit in a three-dimensional Galactic potential, which consists of a point-mass bulge, a \citet{miyamoto75} disk (with $a=4.5\,$kpc and $b=0.5\,$kpc), and a logarithmic halo potential. The combined mass profiles of all three potentials result in a circular velocity of 220 km/s at a galactocentric distance of $8.5\,$kpc. The bulge and disk have masses of $1.5 \times 10^{10}$ and $5 \times 10^{10} M_{\odot}$ respectively \citep{xue08}. \citet{aarseth03} and \citet{praagman10} describe the incorporation of the Galactic potential into NBODY6. All models were set to orbit in the plane of the disk such that a cluster on a circular orbit experiences a static tidal field, and will not be subject to factors such as tidal heating or disk shocking.

For the purposes of our study, the only parameters which are important and change from model to model are initial cluster half-mass radius, initial cluster position, and initial cluster velocity which determine the shape of the orbit. Our first models were for clusters with orbital eccentricities of 0.25, 0.5, 0.75, and 0.9, where eccentricity is defined as $e = \frac{R_{a}-R_p}{R_{a}+R_p}$. All of these have the same perigalactic distances of $6\,$kpc and initial half-mass radii $r_{m,i}$ of 6 pc. These clusters are located at perigalacticon at time zero. For comparison purposes a model was simulated with a circular orbit at perigalacticon (6 kpc) and four more with circular orbits at the apogalactic distance of each eccentric cluster ($10\,$kpc, $18\,$kpc, $43\,$kpc, and $104\,$kpc). These simulations allow us to directly compare the properties of a globular cluster on an eccentric orbit to clusters on circular orbits at both perigalacticon and apogalacticon. 

Our set of models also included re-simulations of the cluster with an orbital eccentricity of 0.5, and the corresponding e = 0 perigalactic and apogalactic simulations, but with initial half-mass radii of $4\,$pc, $2\,$pc, $1\,$pc, and $0.5\,$pc. This range allows us to study the influence of initial cluster half-mass radius. Both sets of models are summarized in Table \ref{table:modparam}. Model names are based on orbital eccentricity (e.g. e05), circular radius at apogalacticon (e.g. r18), and initial half mass radius (rm6). Hence a model cluster with a perigalactic distance of $6\,$kpc, apogalactic distance of $18\,$kpc (orbital eccentricity of 0.5), and an initial half-mass radius of $6\,$pc would be labeled e05r18rm6.

\begin{table}
 \caption{Model Input Parameters}
  \label{table:modparam}
  \begin{center}
    \begin{tabular}{lcccc}
      \hline\hline
      {Model Name} & {$r_{m,i}$} & {$R_p$} & {$v_p$} & {e} \\
      { } & {pc} & {kpc} & {km/s} { } \\
      \hline

e0r6rm6 & 6 & 6 & 212 & 0 \\
e025r10rm6 & 6 & 6 & 280 & 0.25 \\
e0r10rm6 & 6 & 10 & 224.5 & 0 \\
e05r18rm6 & 6 & 6 & 351.5 & 0.5 \\
e0r18rm6 & 6 & 18 & 232 & 0 \\
e075r43rm6 & 6 & 6 & 455 & 0.75 \\
e0r43rm6 & 6 & 43 & 229.95 & 0 \\
e09r104rm6 & 6 & 6 & 543.5 & 0.9 \\
e0r104rm6 & 6 & 104 & 225.25 & 0 \\

\hline

e0r6rm4 & 4 & 6 & 212 & 0 \\
e05r18rm4 & 4& 6 & 351.5 & 0.5 \\
e0r18rm4 & 4 & 18 & 232 & 0 \\
e0r6rm2 & 2 & 6 & 212 & 0 \\
e05r18rm2 & 2& 6 & 351.5 & 0.5 \\
e0r18rm2 & 2 & 18 & 232 & 0 \\
e0r6rm1 & 1 & 6 & 212 & 0 \\
e05r18rm1 & 1& 6 & 351.5 & 0.5 \\
e0r18rm1 & 1 & 18 & 232 & 0 \\
e0r6rm05 & 0.5 & 6 & 212 & 0 \\
e05r18rm05 & 0.5 & 6 & 351.5 & 0.5 \\
e0r18rm05 & 0.5 & 18 & 232 & 0 \\
      \hline\hline
    \end{tabular}
  \end{center}
\end{table}

\section{Influence of Orbital Eccentricity \label{sthree}}

The first portion of this study will focus solely on models with an initial $r_m$ equal to $6\,$pc, with the only difference between each model being their Galactic orbits. However, we first need to determine whether any given star is bound to the cluster. In a cluster-centric coordinate system, we define the x-axis as pointing away from the galactic center, the y-axis pointing in the direction of motion of the cluster, and the z-axis pointing perpendicular to the orbital plane. In this coordinate system the energy of an individual star can be written as:

\begin{equation}\label{energy}
E = \frac{1}{2} (\dot{x}^2 + \dot{y}^2 + \dot{z}^2) - \sum_{i=1}^{N-1} \frac{ G m_i}{\| r-r_i \|} - \frac{1}{2} \Omega^2 (z^2 - \upsilon x^2)
\end{equation}

\noindent where the second term is the potential energy due to the remaining N-1 stars in the simulation, each with mass $m_i$ and located a distance $r_i$ from the star. The third term is the tidal potential with $\Omega^2$ equal to the orbital frequency of the cluster. Here $\upsilon$ is a dimensionless positive coefficient defined below in Equation \ref{upsilon}. The tidal potential, taken from \citet{bertin08}, results in a stretching of the cluster in the x-direction, no change in the y-direction, and a compression in the z-direction. If the resultant energy is less than zero the star is bound, otherwise it is considered unbound.

We considered additional criteria for determining whether a star is bound or unbound in addition to the energy calculation. Other studies have invoked a distance cutoff such that the stars' cluster-centric distance must be greater than the cluster perigalactic or instantaneous tidal radius for it to be unbound \citep[e.g.][]{takahashi12}. It has also been suggested that a star's velocity plays a role in whether or not it can be considered unbound \citep[e.g.][]{kupper10, kupper12}. However, we found these additional criteria did not change any of the results found in Section 3 as they only effected a small percentage of simulated stars. Therefore, we only require that a star's energy as given by Equation \ref{energy} to be greater than zero for the star to be considered unbound.

\begin{figure}[tbp]
\centering
\includegraphics[width=\columnwidth]{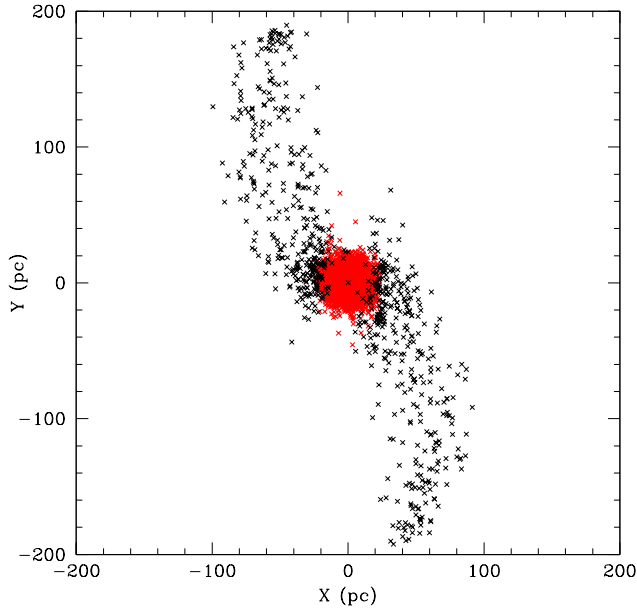}
\caption{Snapshot of a model cluster on a circular orbit at 6 kpc after 32 orbits (5680 Myr). Bound stars are marked in red.}
  \label{fig:xy300}
\end{figure}

Figure \ref{fig:xy300} shows a model cluster at a representative timestep. The tidal tails formed by escaping stars are clearly visible in our simulations. The densely populated spherical collection of stars marked in red are those that satisfy our boundedness criterion and are considered cluster members. The unbound stars that appear close to the centre of the cluster are foreground stars with a large z coordinate and are simply projected onto the cluster in the x-y plane. These tails have no effect on our determination of theoretical tidal radii or observed limiting radii as we only consider stars that are bound to the cluster.

\subsection{Mass}

\begin{figure}[tbp]
\centering
\includegraphics[width=\columnwidth]{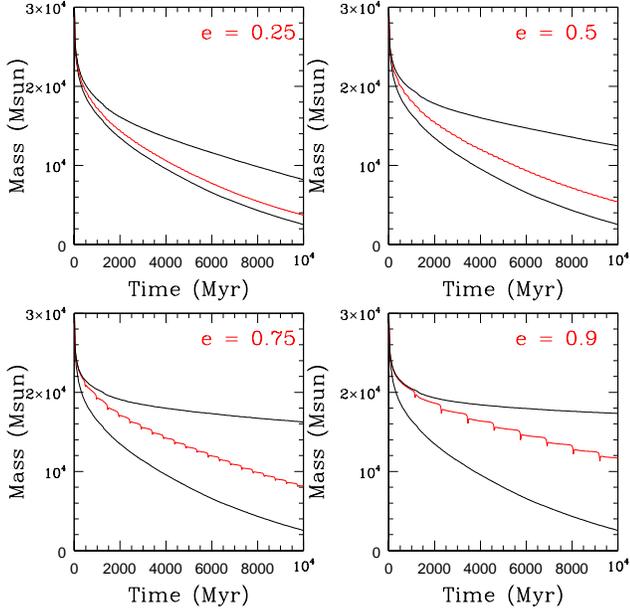}
\caption{The evolution of total cluster mass over time. The red lines correspond to models with orbital eccentricities as labelled in each panel. In each plot, the lower black line corresponds to a cluster with a circular orbit at perigalacticon ($6\,$kpc), while the upper black line corresponds to a cluster with a circular orbit at apogalacticon.}
  \label{fig:m_t_b}
\end{figure}

We first wish to study how orbital eccentricity influences the total mass, or more specifically the mass loss, of a star cluster over time, which then plays an important role in determining tidal radii ($r_t \propto M^\frac{1}{3}$). The total mass of stars bound to the cluster in each model is illustrated in Figure \ref{fig:m_t_b}. In general, mass loss is due to stellar evolution, evaporation due to two-body interactions, and tidal stripping. For clusters with circular orbits, in all cases the apogalactic cluster loses mass at a lower rate than the perigalactic case as a result of less tidal stripping.

The e=0.25 case loses mass at almost the same rate as if it had a circular orbit at its perigalactic distance, but the final mass is still notably larger than the perigalactic case. Since eccentric clusters spend the majority of their time away from perigalacticon, they too will be subject to less tidal stripping than an ideal cluster that spends all its time at perigalacticon. As eccentricity increases, the mass-loss profile shifts further away from the perigalactic case and closer to a cluster with a circular orbit at apogalacticon.

At higher eccentricities the mass no longer smoothly decreases, in contrast to the circular orbit cases. Instead periodic fluctuations are present. The minima of these fluctuations correspond to perigalactic passes, where the rapid increase in tidal field strength results in episodes of significant mass loss. These fluctuations suggest that a greater change in tidal field strength between apogalacticon and perigalacticon results in stars gaining more energy at or near perigalacticon. 

Especially interesting in the lower right panel of Figure \ref{fig:m_t_b} is the fact that once a cluster undergoes significant mass loss during a perigalactic pass, the cluster starts to regain mass before resuming its mean mass loss rate. In these intervals just after perigalacticon, the cluster is re-capturing some of the stars which were previously unbound. Stated differently, many of the stars that were formally unbound at $R_p$ drift away slowly enough that they are recaptured as the cluster moves back outward and its instantaneous tidal radius expands again. Furthermore, these fluctuations suggest that while a perigalactic pass does have a strong effect on an eccentric cluster, the cluster cannot be treated as if it had a circular orbit at $R_p$. These results are in agreement with the findings of \citet{kupper12} discussed earlier.

\subsection{Half-mass Radius}

We next consider how orbital eccentricity can influence the half-mass radius $r_m$ of bound stars within a globular cluster. It should be noted that $r_m$ is not the same as the \textit{half-light radius} $r_h$ (also known as the effective radius), which is a directly observable parameter. In our simulations, the half-mass radius is always slightly larger than the half-light radius. 

\begin{figure}[tbp]
\centering
\includegraphics[width=\columnwidth]{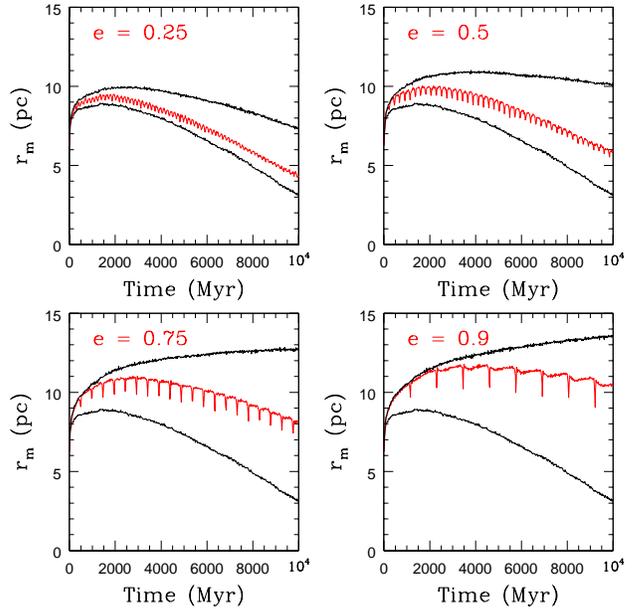}
\caption{The evolution of half mass radius over time. The red lines correspond to models with orbital eccentricities as labelled in each panel. In each plot, the lower black line corresponds to a cluster with a circular orbit at perigalacticon ($6\,$kpc), while the upper black line corresponds to a cluster with a circular orbit at apogalacticon.}
  \label{fig:rm_t_b}
\end{figure}

The results of our simulations are illustrated in Figure \ref{fig:rm_t_b}. The initial increase in $r_m$ during the first $\sim 2000$ Myr in all cases is driven by two-body relaxation and stellar evolution mass-loss. However once the cluster is relaxed, tidal stripping becomes the dominant dynamical process.

The $r_m$ profiles of the apogalactic cases in the lower panels do not begin to decrease after 2000 Myr, but instead continue to increase up to 10 Gyr. As discussed in the next section this trend is due to the fact that these clusters are barely tidally filling, so can still expand and not be subject to tidal stripping.

Similar to the results of Section 3.1, for low eccentricities the $r_m$ profile of the eccentric model cluster is comparable to the circular orbit case at perigalacticon. Increasing eccentricity brings the $r_m$ profile closer to the apogalactic case on the average, but increasing eccentricity again results in sharp fluctuations in the $r_m$ profile which correspond to perigalactic passes. The trends shown in Figure \ref{fig:rm_t_b} reveal what is perhaps the most striking difference between clusters on static circular orbits (the classically assumed case) and ones on more realistic eccentric orbits. If cluster limiting radii are imposed at perigalacticon, we would expect the minima of the eccentric $r_m$ profile to be equal to the $r_m$ profile of a cluster orbiting at $R_p$. While a high-e cluster may briefly expand after a perigalactic pass, the next perigalactic pass would restore $r_m$ to a size equal to the perigalactic case. But not even at perigalacticon does the $r_m$ of the eccentric cluster equal the perigalactic case. Instead, the time averaged $r_m$ is reflective of a tidal field weaker than the field at perigalacticon, in agreement with \citet{kupper10}.

After a perigalactic pass, Figure \ref{fig:rm_t_b} illustrates again that the cluster is able to increase in size. Especially apparent in the lower right panel of Figure \ref{fig:rm_t_b} is the fact that the cluster is able to increase to a size greater than its mean $r_m$. Inspection of our $N$-body models shows that this increase in size is due to a combination of re-capturing some of the previously unbound stars \citep{kupper12} and the stars in the inner region of the cluster gaining enough energy to move outward and repopulate the halo of the cluster. These statements are discussed in further detail in Section 5.0.

\subsection{Tidal and Limiting Radii}

The Jacobi radius represents a theoretical surface around a cluster past which a star cannot pass and still remain bound. The Jacobi radius allows for the calculation of the instantaneous tidal radius of each model cluster as a function of time. To calculate the instantaneous tidal radius of each model cluster, we require a derivation of cluster tidal radius which takes into consideration the tidal field of the host galaxy. Assuming only that the tidal field must be spherically symmetric, the theoretical tidal radius as derived by \cite{bertin08} is:

\begin{equation} \label{rt}
r_t=(\frac{GM}{\Omega^2\upsilon})^{1/3}
\end{equation}

\noindent where $\Omega$, $\kappa$ and $\upsilon$ are defined as:

\begin{equation}
\Omega^2=(d\Phi_G(R)/dR)_{R_{p}}/R_{p}
\end{equation}
\begin{equation}
\kappa^2=3\Omega^2+(d^2\Phi_G(R)/dR^2)_{R_{p}}
\end{equation}
\begin{equation}\label{upsilon}
\upsilon=4-\kappa^2/\Omega^2
\end{equation}

\begin{figure}[tbp]
\centering
\includegraphics[width=\columnwidth]{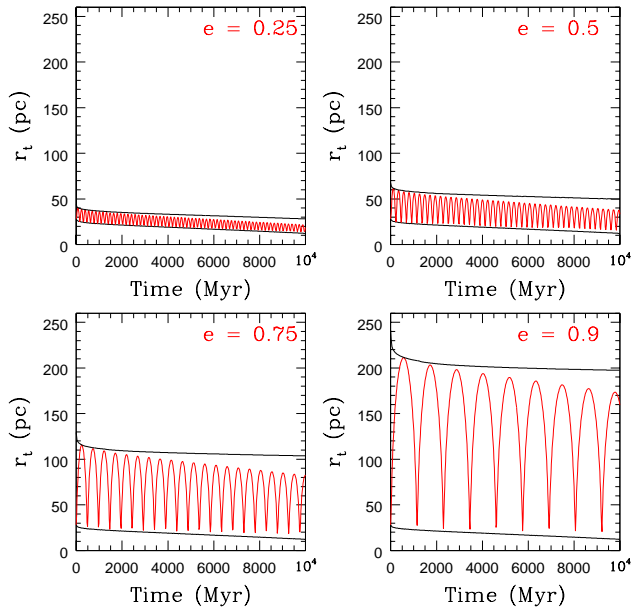}
\caption{The evolution of the instantaneous tidal radius over time. The red lines correspond to models with orbital eccentricities as labelled in each panel. In each plot, the lower black line corresponds to a cluster with a circular orbit at perigalacticon ($6\,$kpc), while the upper black line corresponds to a cluster with a circular orbit at apogalacticon.}
  \label{fig:rt_t_b}
\end{figure}

\noindent $\Phi_G$ is the galactic potential, $R_{p}$ is the perigalactic distance, $\Omega$ is the orbital frequency of the cluster, $\kappa$ is the epicyclic frequency of the cluster at $R_p$, and $\upsilon$ is a positive dimensionless coefficient. Using the tidal field of the Milky Way discussed in Section 2.0 and the mass and galactocentric distance of the model clusters at each time step, we calculate the instantaneous tidal radius of each model. The results of these calculations are shown in Figure \ref{fig:rt_t_b}. Here the instantaneous $r_t$ increases and decreases periodically along the orbit, but never quite reaches the perigalactic and apogalactic cases due to differences in mass loss rates among all three cases.

Next we compare the tidal radius to the limiting radius. For a simulated cluster, since we know which stars are bound or unbound, we could call the limiting radius of the cluster the distance to the farthest bound star, but this approach introduces some significant problems. First, since cluster tidal radii are calculated for stars with \textit{circular prograde orbits}, any star with a retrograde and/or eccentric orbit within its cluster can remain bound beyond the nominal tidal radius \citep{read06}. Second, any star in the process of escaping the cluster can reach large clustercentric distances before becoming energetically unbound. Third, the stars along the y-axis of the cluster (the direction of motion) are unaffected by the tidal potential in Equation \ref{energy}, allowing them to also remain bound at larger clustercentric distances. These three issues cause the true limiting radius of the cluster to change dramatically from time-step to time-step. To gain a more stable indication of cluster size, we instead focus on the x-axis of the cluster, the axis along which the tidal radius is calculated, and define the limiting radius as the average x-coordinate of all stars with $\|x\| > r_t$. This calculation typically involves less than $1 \%$ of the total cluster population. While this is not the true limiting radius of the cluster and will always be slightly larger than the true tidal radius, it acts as a tracer of the outer region that is less affected by individual extreme outliers. If a cluster is tidally over-filling, the limiting radius will still be significantly larger than the tidal radius. For a cluster that is tidally under-filling, the limiting radius is simply the distance to the outermost bound star.

\begin{figure}[tbp]
\centering
\includegraphics[width=\columnwidth]{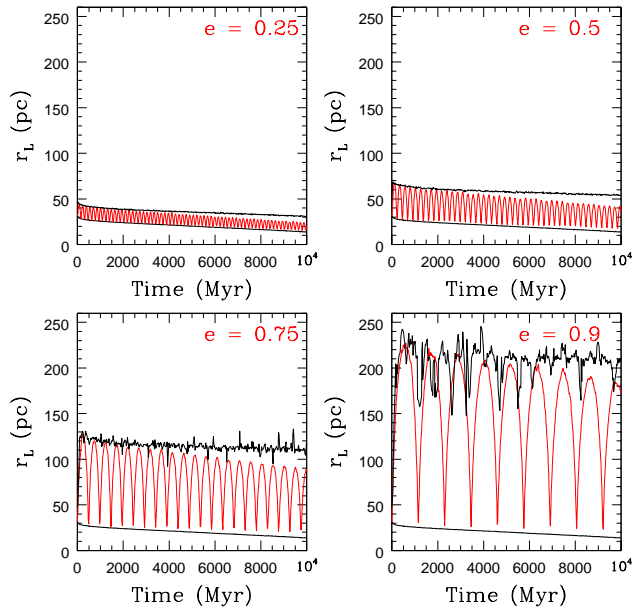}
\caption{The evolution of the limiting radius over time. The red lines correspond to models with orbital eccentricities as labelled in each panel. In each plot, the lower black line corresponds to a cluster with a circular orbit at perigalacticon ($6\,$kpc), while the upper black line corresponds to a cluster with a circular orbit at apogalacticon.}
  \label{fig:rl_t_b}
\end{figure}

In Figure \ref{fig:rl_t_b} we show this empirically determined $r_L$ for each model as a function of time. For circular orbits, on average the limiting radius of the cluster decreases smoothly as a result of mass loss. For eccentric orbits, the small fluctuations with perigalactic passes in Figures \ref{fig:m_t_b} and \ref{fig:rm_t_b} are much more prevalent in Figure \ref{fig:rl_t_b}. Comparing Figure \ref{fig:rt_t_b} to Figure \ref{fig:rl_t_b}, the fluctuations in both figures indicate that the limiting radius behaves the same as the instantaneous tidal radius.

It should be noted that the apogalactic cases for e = 0.75 and 0.9 in Figure \ref{fig:rl_t_b} are not smooth due to the fact that these clusters are barely tidally filling and their limiting radii are easily influenced by individual escaping stars.

Directly comparing limiting radii in Figure \ref{fig:rl_t_b} and tidal radii in Figure \ref{fig:rt_t_b}, all circular orbits have a relatively constant ratio at approximately $\frac{r_L}{r_t} = 1.1$. Since we expect $r_L$ to slightly overestimate $r_t$, a ratio of 1.1 suggests that these clusters are approximately tidally filling. For eccentric clusters the ratio is in general also 1.1, suggesting the clusters come close to filling their instantaneous tidal radius at all times. Fluctuations in the ratio for the e = 0.75 and 0.9 cases indicate that after a perigalactic pass the cluster is slightly tidally under-filling and works to fill its instantaneous tidal radius on the way to apogalacticon. When travelling back in from apogalacticon to perigalacticon, the cluster will remain tidally filled and lose stars to tidal stripping as the instantaneous tidal radius shrinks.

\section{Influence of Initial Cluster Half-Mass Radius \label{sfour}}

\begin{figure}[tbp]
\centering
\includegraphics[width=\columnwidth]{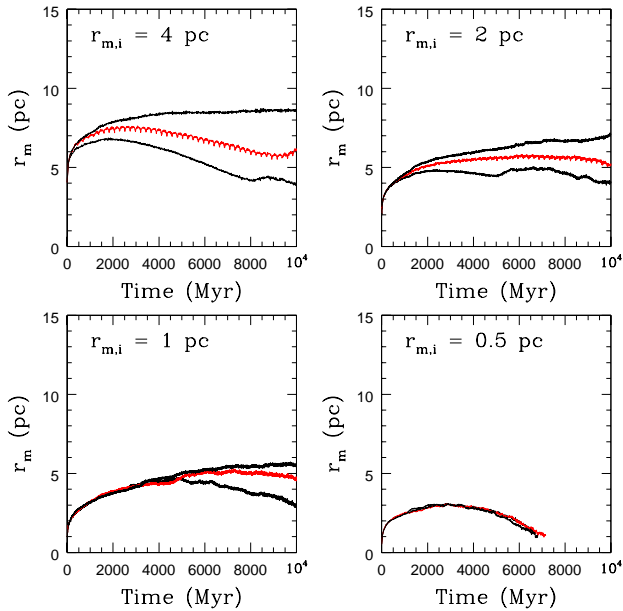}
\caption{The evolution of half mass radius over time. The upper left, upper right, lower left, and lower right panels are for simulations with initial half mass radii of $4\,$pc, $2\,$pc, $1\,$pc, and $0.5\,$pc respectively. The red lines correspond to models with orbital eccentricities of 0.5, while the lower black lines correspond to a cluster with a circular orbit at perigalacticon ($6\,$kpc) and the upper black lines correspond to a cluster with a circular orbit at apogalacticon ($18\,$kpc).}
  \label{fig:rm_t_rmi}
\end{figure}

Up until this point we have only considered clusters with initial half-mass radii of $6\,$pc. This initial half-mass radius was chosen simply to ensure that the model clusters with $R_{p} = 6\,$kpc would be tidally filling. As seen in Figure \ref{fig:rm_t_b} this produces clusters with sizes at 10 Gyr ranging from 3 to $14\,$pc, which are larger than most (but not all) real globular clusters. In an attempt to produce Milky Way-like clusters which have a mean effective radius of $2.5\,$pc, we re-simulated the e0r6rm6, e05r18rm6, and e0r18rm6 models with initial half-mass radii of $4\,$pc, $2\,$pc, $1\,$pc, and $0.5\,$pc. The results are shown in plots of half mass radius versus time in Figure \ref{fig:rm_t_rmi}.

The rm4 clusters closely resemble the rm6 clusters, with similar periodic fluctuations with perigalactic passes and the final $r_m$ values for the perigalactic, eccentric, and apogalactic cases. However, the rm4 clusters undergo smaller initial expansion due to two-body interactions than the rm6 clusters, and thus when outer region stars are being removed through tidal stripping, since the majority of the mass is concentrated in the inner region, the mass-loss profile is less affected.

This issue of initial size becomes even more significant in the rm2, rm1, and rm05 cases. The periodic fluctuations on the eccentric orbit are barely visible in the rm2 cluster, and non-existent in the rm1 and rm05 clusters. The rm1 cases are significantly smaller at 10 Gyr, approaching the typical $\sim$ 2 - $3\,$pc size that match the majority of real globular clusters. The rm05 models have completely dissolved by 10 Gyr.

These small clusters are only tidally filling in the sense that two-body interactions have pushed \textit{some} bound stars to orbits that take them out to the instantaneous tidal radius. With the majority of the bound stars located in the inner regions of the cluster, tidal stripping is not the dominant form of mass loss and the influence of the galactic potential and cluster orbit are minimized. Instead, stellar evolution and two-body interactions are the dominant forms of mass loss. These clusters would be classified as ``tidally unaffected" \citep{carballo12}. For the rm1 case, the $r_m$ profiles of the perigalactic, eccentric, and apogalactic cases begin to split only after $\sim 5$ Gyr, when stars have been finally pushed to the outer regions of the cluster and tidal stripping is beginning to play an important role. While this is true, it is not due to two-body interactions but instead a result of core collapse. For the rm05 case, not even core collapse can push stars to the outer region of the cluster in order for tidal stripping to occur. In fact, the rm05 clusters all have the same $r_m$ up until the complete evaporation of the cluster at approximately 7 Gyr.

Producing Milky Way-like globular cluster effective radii of 2 - $3\,$pc for clusters on circular orbits at $6\,$kpc or greater appears to require initial $r_m$ size less than $1\,$pc. The $N$-body models reveal that either clusters originally are extremely compact and tidally unaffected, or present-day cluster orbits have changed significantly from the orbit along which the clusters originally formed. Some observational support for this view can be found in recent measurements of very young, massive clusters \citep[e.g.][]{bastian08, bastian12, portegies10, marks10}. However, recent $N$-body simulations by \citet{sippel12} showed that $r_h$ and $r_m$ can be very different because of stellar mass segregation, and produce clusters with final $r_h$ values near $3\,$pc despite large $r_{m,i}$. These issues will be explored in future studies.

\section{Discussion}

\begin{figure}[htbp]
\centering
\includegraphics[width=\columnwidth]{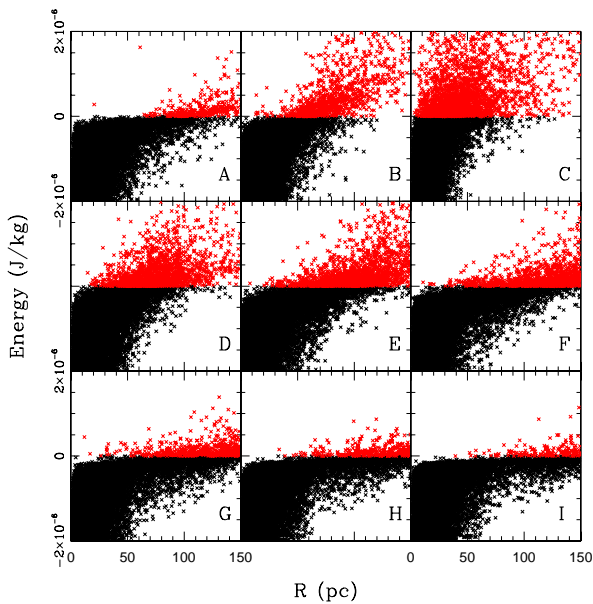}
\caption{The radial distance and energy of stars within model cluster e09r104rm6 for different time steps. Beginning in Panels A and B the cluster is travelling towards perigalacticon. In Panel C the cluster has just left perigalacticon. In Panels D to I the cluster is moving away from perigalacticon. Bound stars are marked as black and unbound stars are marked as red. }
  \label{fig:e_r_t}
\end{figure}

\begin{figure}[tbp]
\centering
\includegraphics[width=\columnwidth]{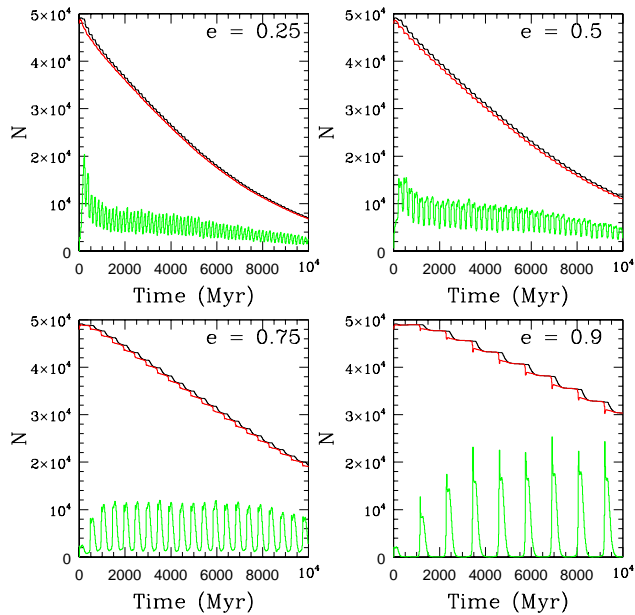}
\caption{The total number of bound (red) and unbound (green) stars in a cluster as a function of time. The black line corresponds to the total number of stars and the orbital eccentricity of the model is labelled in each panel. For comparison purposes, the number of unbound stars has been increased by a factor of 10.}
  \label{fig:n_t_b}
\end{figure}

\begin{figure}[tbp]
\centering
\includegraphics[width=\columnwidth]{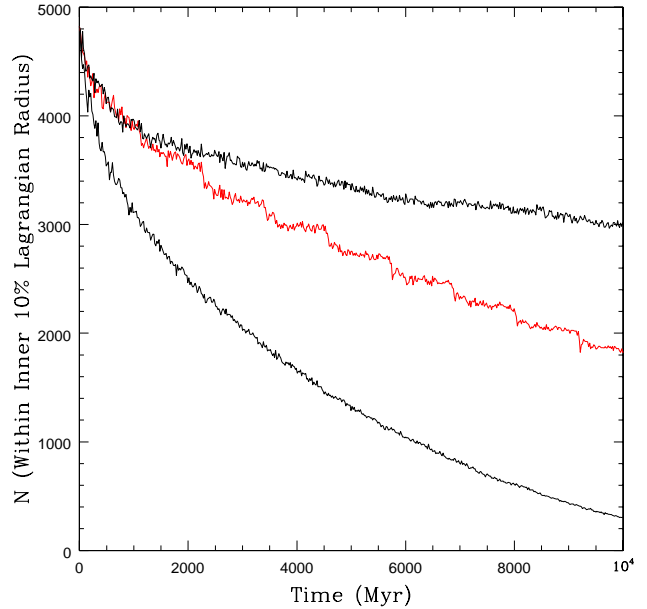}
\caption{The evolution of the number of stars within the inner 10 $\%$ Lagrangian radius over time. The red line correspond to the e=0.9 model (e09r104b). The lower black line corresponds to a cluster with a circular orbit at perigalacticon ($6\,$kpc), while the upper black line corresponds to a cluster with a circular orbit at apogalacticon ($104\,$kpc). The number of stars within the inner 10 $\%$ Lagrangian radius will naturally decrease over time due to two-body encounters, however the eccentric case (red line) illustrates that with each perigalactic pass a significant number of stars move beyond the inner 10 $\%$ Lagrangian radius.}
  \label{fig:nL_t_b}
\end{figure}

A perigalactic pass has three effects on a globular cluster, which we illustrate in Figure \ref{fig:e_r_t} for the e=0.9 model e09r104rm6. In this figure, we plot the energy per unit mass of individual stars (as per Equation \ref{energy}) as a function of radial distance from the cluster center at 9 points in the orbit. Beginning in Panel A of Figure \ref{fig:e_r_t}, for a given time between apogalacticon and perigalacticon there are a few stars that are within close proximity of the cluster but remain unbound (marked in red). As the cluster moves towards $R_p$ and the instantaneous tidal radius shrinks (Panel B), more and more stars become temporarily unbound. As predicted in Section 3.0, even stars in the inner region of the cluster are provided with a significant increase in energy by the tidal shock and can become unbound. Just after the cluster reaches perigalacticon (Panel C), a large number of stars are no longer bound to the cluster. As the cluster moves away from perigalacticon (Panels D to I): 

\begin{itemize}
\item some stars that became unbound escape the cluster (which causes the initial decrease in $r_m$ and $r_L$);
\item some of the stars that are unbound in Panel C return to energies below zero and are recaptured (see Figure \ref{fig:n_t_b});
\item the tidal shock gives stars initially found in the inner region enough additional energy to move outward and fill the orbits vacated by stars which permanently escaped the cluster (see Figure \ref{fig:nL_t_b}). 
\end{itemize}

It is even possible for inner region stars to become temporarily or permanently unbound if they move outward at a rate faster than the instantaneous tidal radius increases.

\section{Predicting Cluster Limiting Radii}

Now that we have shown that limiting radii are not imposed at perigalacticon, it is useful to know how to calculate a meaningful number that predicts the limiting radius of a globular cluster on an eccentric orbit. As we saw in Figures \ref{fig:rt_t_b} and \ref{fig:rl_t_b}, the limiting radius essentially traces the instantaneous tidal radius. However the ratio between cluster limiting radius and instantaneous tidal radius undergoes small periodic variations as a function of the location of a cluster along its orbit.

\begin{figure}[htbp]
\centering
\includegraphics[width=\columnwidth]{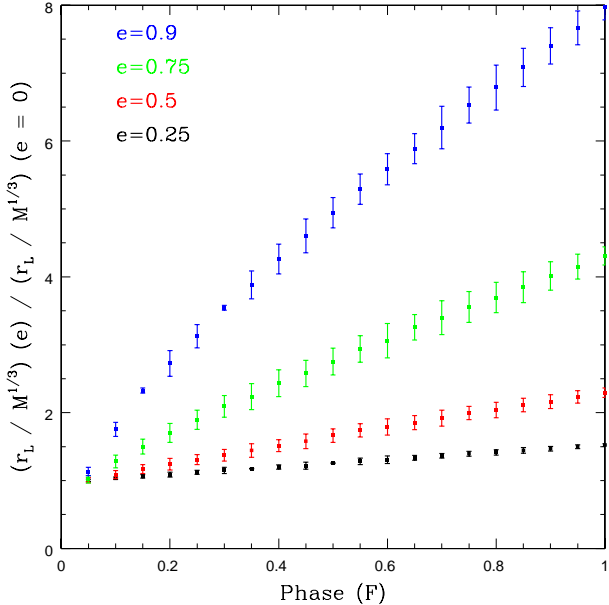}
\caption{The ratio of the mass normalized limiting radius of model clusters with eccentric orbits to the mass normalized limiting radius of a cluster with a circular orbit at perigalacticon as a function of orbital phase as defined in Equation \ref{phase}. Error bars represent an uncertainty of $1 \sigma$.}
  \label{fig:rl_rat_phase}
\end{figure}

Orbital phase is defined as:

\begin{equation}\label{phase}
F = \frac{R_{gc} - R_p}{R_{a}-R_p}
\end{equation}

\noindent such that the cluster has $F = 0 $ at perigalacticon and $F = 1$ at apogalacticon. A cluster with a circular orbit will always have $F = 0$. The median limiting radius of the eccentric cluster as a function of orbital phase is then determined in order to calculate the ratio of the instantaneous limiting radius to the limiting radius of the e = 0 perigalactic case ($\frac{r_L(e)}{r_L(e = 0)}$), both normalized by mass. This ratio is plotted as a function of phase in Figure \ref{fig:rl_rat_phase}. It is important to note that we have ignored the first 2000 Myr of evolution for each model cluster when evaporation due to two-body relaxation is the dominant source of mass loss.

For a given orbital eccentricity, $\frac{r_L(e)}{r_L(e = 0)}$ changes almost linearly with phase F. It is interesting to note that we observed a second order effect that the rate at which the cluster expands is lower than the rate at which it contracts. When the cluster is moving away from perigalacticon, it works to fill its expanding tidal radius. Conversely a cluster moving towards perigalacticon would be larger as it is always tidally filling on the way inward.
 
The \textit{rate} at which the limiting radius of the cluster increases and decreases as a function of orbital phase is a strong function of orbital eccentricity. These rates were determined explicitly by finding the slopes of each line, where the y-intercept is forced to equal one. The slopes are listed in Table \ref{table:slopes}, along with the associated uncertainty ($1 \sigma$).

\begin{table}
  \caption{Lines of Best Fit}
  \label{table:slopes}
  \begin{center}
    \begin{tabular}{lcccc}
      \hline\hline
      {Orbital Eccentricity} & {Slope} & {Uncertainty}  \\
      \hline

0.25 & 0.512 & 0.007 \\
0.5 & 1.29 & 0.04 \\
0.75 & 3.37 & 0.07 \\
0.9 & 7.84 & 0.07 \\
      \hline\hline
    \end{tabular}
  \end{center}
\end{table}

A smooth relationship between slope and eccentricity emerges, that can be fit with an exponential, and allows us to propose a purely analytical correction to the calculation of the tidal radius of a globular cluster. For a globular cluster on an orbit with eccentricity E, with a tidal radius at perigalacticon $r_t(R_p)$ and located at a phase F in its orbit, its limiting radius is equal to

\begin{equation}\label{correct}
r_L(F) = r_t(R_p) (1 + a \ F \ e^{b \ E})
\end{equation}

where $a = 0.17 \pm 0.03$ and $b=4.1 \pm 0.2$. Note that since this calculation involves a single cluster over the course of a single orbit, the mass normalization is no longer necessary as cluster mass will not have changed significantly over a fraction of one orbit.

The next step is to simulate a larger suite of model clusters ranging in initial cluster mass and half-mass radii to determine if these parameters play a role in the correction factor suggested above. However, regardless of the influence of cluster mass or initial half-mass radius, all tidally affected simulations follow the rule that the limiting radius traces the instantaneous tidal radius rather well. Thus if full orbital information or phase F is unknown, the calculation of the instantaneous tidal radius is a reasonable estimate of the limiting radius of a cluster. For globular clusters in other galaxies, in which only their projected galactocentric distances are known, it may be possible to determine their theoretical tidal radius based on their present King radius $r_k$. Future work will explore this possibility.

\subsection{Application to the Milky Way}

\begin{figure}[tbp]
\centering
\includegraphics[width=\columnwidth]{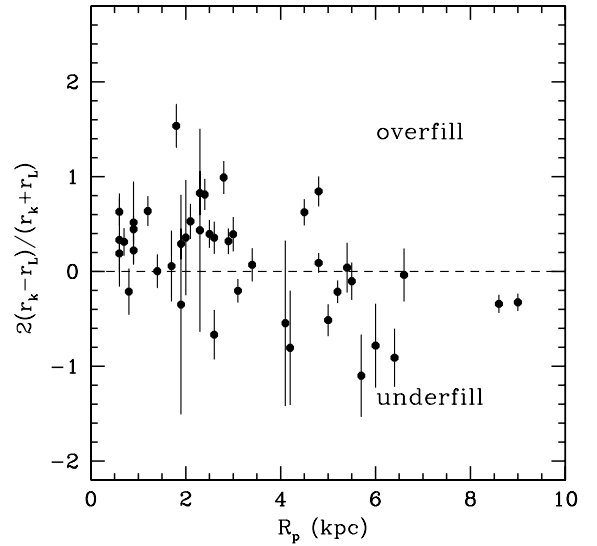}
\caption{Ratio of difference between fitted \citet{king62} radius ($r_k$) and limiting radius ($r_L$) to the average of the two radii versus perigalactic distance for Galactic globular clusters. Limiting radii have been calculated based on the orbital eccentricity and phase of a cluster as given by Equation \ref{correct}.}
  \label{fig:rt_rgc_norm_corr}
\end{figure}

For many Milky Way globular clusters, their current galactocentric distance, orbital eccentricity and orbital phase are known. In Figure \ref{fig:rt_rgc_norm_corr} the revised, fully corrected version of Figure \ref{fig:rt_rp_norm} is illustrated, where $r_L$ is now the phase-corrected value from Equation \ref{correct}. We now see more tidally under-filling clusters and the scatter of points more nearly around zero. A stronger agreement between theory and observations emerges. Correcting for using a non-spherically symmetric potential in calculating tidal radii and improved methods for determining observational limiting radii will likely strengthen this comparison further.

\section{Conclusions and Future Work \label{sfive}}

Globular clusters have been simulated with a range of both circular and eccentric orbits. After determining which stars are bound to the cluster at a given time, we show that while eccentric clusters undergo episodes of significant mass loss during a perigalactic pass, their time averaged mass loss rate reflects a tidal field less than the tidal field at perigalacticon. Additionally it was found that clusters are able to re-capture unbound stars after a perigalactic pass as their instantaneous tidal radius increases. 

Second, we show that the half-mass radius of a globular cluster increases and decreases about a mean value over the course of an orbit. These fluctuations suggest that the perigalactic pass also has the effect of energizing inner region stars to larger orbits. Finally, we find that the limiting radius of a cluster traces its instantaneous tidal radius at all times.

These findings argue against the historical assumption that globular cluster tidal radii, and by extension limiting radii, are imposed at perigalacticon for clusters that do not have circular orbits. While it remains true that the half-mass relaxation time is greater than one orbital period, the cluster does not need to fully relax in order to expand. The eccentric orbit introduces an effect of tidal shocking that is not experienced by clusters in a static potential (circular orbit).

While the instantaneous tidal radius is a useful first approximation of the limiting radius, we have proposed an analytically determined correction factor that is a function of orbital eccentricity and phase. This correction leads to a much stronger agreement between the predicted limiting radii and observational \cite{king62} radii of Milky Way globular clusters. Future studies will explore how the correction factor depends on initial mass or initial half-mass radius and how corrected limiting radii are related to King radii.

Since the tidal field of the Milky Way is not spherically symmetric, correcting limiting radii based on eccentricity and orbital phase is not the final step. We still need to correct for orbital inclination to account for factors like disk shocking and tidal heating, which may reveal important effects for the Milky Way and other disk galaxies. However, the present results already have clear applicability to elliptical galaxies, which have more nearly spherical potentials. We are currently investigating $N$-body simulations in these directions. The ultimate goal is to be able to predict the limiting radius of any tidally affected globular cluster, given its orbit, galactocentric position and the galactic potential of the host galaxy.

\section{Acknowledgements}

JW would like to acknowledge funding through the A. Boyd McLay Ontario Graduate Scholarship. JW would also like to thank Juan P. Madrid and Anna Sippel for valuable discussions and correspondence. AS and WEH acknowledge financial support through research grants from the Natural Sciences and Engineering Research Council of Canada.

%************************************BIBLIOGRAPHY************************************

\end{document}